\documentclass{PoS}

\title{Nonperturbative calculations in light-front QED}

\ShortTitle{Nonperturbative calculations in light-front QED}

\author{\speaker{Sophia CHABYSHEVA}\\
        University of Minnesota-Duluth, USA\\
        E-mail: \email{sophia@mail.physics.smu.edu}}

\abstract{%
The methods of light-front quantization and Pauli--Villars
regularization are applied to a nonperturbative calculation
of the dressed-electron state in quantum electrodynamics.
This is intended as a test of the methods in a gauge theory,
as a precursor to possible methods for the nonperturbative
solution of quantum chromodynamics.  The electron state
is truncated to include at most two photons and no positrons
in the Fock basis, and the wave functions of the dressed
state are used to compute the electrons's anomalous
magnetic moment.  A choice of regularization that preserves
the chiral symmetry of the massless limit is critical for the
success of the calculation.
}

\FullConference{Light Cone 2010 - LC2010\\
		June 14-18, 2010\\
		Valencia, Spain}

\def\bea{\begin{eqnarray}}
\def\eea{\end{eqnarray}}
\def\be{\begin{equation}}
\def\ee{\end{equation}}
\newcommand{\ub}[1]{\underline{#1}}

\begin{document}

\section{Introduction}

The purpose of this work
is to explore a nonperturbative method that
can be used to solve for the bound states of quantum field theories,
in particular QCD.  The problem is notoriously difficult,
and there are only a few approaches.  These include
lattice gauge theory~\cite{Lattice}, 
the transverse lattice~\cite{TransLattice},
Dyson--Schwinger equations~\cite{DSE},
Bethe--Salpeter equation,
similarity transformations combined with construction of
effective fields~\cite{Glazek},
and light-front Hamiltonians with
either standard~\cite{bhm} or sector-dependent
parameterizations~\cite{SectorDependent,Karmanov,Vary}.  We use
the light-front Hamiltonian approach with Pauli--Villars 
(PV)~\cite{PauliVillars} regularizaton and 
standard parameterization, where the bare
parameters of the Lagrangian do not depend
on the Fock sector.  As a test in a gauge
theory, we consider light-front QED and
specifically the eigenstate of the dressed
electron and its anomalous moment~\cite{ChiralLimit,Thesis,SecDep,TwoPhotonQED}.
       
We use light-cone coordinates~\cite{Dirac,DLCQreview}, chosen in order to have 
well-defined Fock-state expansions and a simple vacuum.
The time coordinate is $x^+=t+z$ and the space
coordinates are $\underline{x}=(x^-,\vec{x}_\perp)$, with 
$x^-\equiv t-z$ and $\vec{x}_\perp=(x,y)$.  The light-cone
energy is $p^-=E-p_z$, and the three-momentum is
$\underline{p}=(p^+,\vec{p}_\perp)$, with
$p^+\equiv E+p_z$ and $\vec{p}_\perp=(p_x,p_y)$.
The mass-shell condition
$p^2=m^2$ becomes $p^-=\frac{m^2+p_\perp^2}{p^+}$.
The simple vacuum follows from the positivity of the
plus component of the momentum:
$p^+\equiv \sqrt{m^2+p_z^2+p_\perp^2}+p_z>0$.

To regulate QED, we use the Pauli--Villars technique~\cite{PauliVillars}.           
The basic idea is to subtract from each integral a contribution
of the same form but of a PV particle with a much larger mass.
This can be done by adding negative metric particles to the
Lagrangian.  A particular advantage of
PV regularization is preservation of at least some symmetries;
in particular, it is automatically relativistically covariant.

From the PV-regulated light-front QED Lagrangian, we construct
the Hamiltonian ${\cal P}^-$ and solve the mass eigenvalue
problem ${\cal P}^-|\ub{P}\rangle=\frac{M^2}{P^+}|\ub{P}\rangle$
in the approximation that the electron eigenstate is a truncated
Fock-state expansion with at most two photons and no positrons.
From this approximate eigenstate, we compute the anomalous magnetic
moment, as a test of the method.

\section{Light-front QED}

The light-front QED Lagrangian with one PV fermion and two PV photons is
\bea \label{eq:Lagrangian}
{\cal L} &= & \sum_{i=0}^2 (-1)^i \left[-\frac14 F_i^{\mu \nu} F_{i,\mu \nu} 
         +\frac12 \mu_i^2 A_i^\mu A_{i\mu} 
         -\frac12 \left(\partial^\mu A_{i\mu}\right)^2\right] \\
  && + \sum_{i=0}^1 (-1)^i \bar{\psi_i} (i \gamma^\mu \partial_\mu - m_i) \psi_i 
  - e_0 \bar{\psi}\gamma^\mu \psi A_\mu ,  \nonumber
\eea
with
\be \label{eq:NullFields}
  \psi =  \sum_{i=0}^1 \psi_i, \;\;
  A_\mu  = \sum_{i=0}^2 \sqrt{\xi_i}A_{i\mu}, \;\;
  F_{i\mu \nu} = \partial_\mu A_{i\nu}-\partial_\nu A_{i\mu} .
\ee
The coupling coefficients $\xi_i$ are constrained by
$\xi_0=1$ and $\sum_{i=0}^2(-1)^i\xi_i=0$,
and the requirement of chiral symmetry restoration
in the limit of zero electron mass.  At one loop,
the chiral symmetry constraint becomes~\cite{ChiralLimit}
$\sum_{i=0}^2(-1)^i\xi_i
   \frac{\mu_i^2/m_1^2}{1-\mu_i^2/m_1^2}\ln(\mu_i^2/m_1^2)=0$;
for nonperturbative solutions with more than one
photon in the basis, the constraint must be imposed
numerically~\cite{TwoPhotonQED}.

The light-front Hamiltonian without antifermion terms is
then of the form~\cite{SecDep}
\bea \label{eq:QEDP-}
\lefteqn{{\cal P}^-=
   \sum_{i,s}\int d\ub{p}
      \frac{m_i^2+p_\perp^2}{p^+}(-1)^i
          b_{i,s}^\dagger(\ub{p}) b_{i,s}(\ub{p})} \\
   && +\sum_{l,\mu}\int d\ub{k}
          \frac{\mu_l^2+k_\perp^2}{k^+}(-1)^l\epsilon^\mu
             a_{l\mu}^\dagger(\ub{k}) a_{l\mu}(\ub{k})
          \nonumber \\
   && +\sum_{i,j,l,s,\mu}\int d\ub{p} d\ub{q}\left\{
      b_{i,s}^\dagger(\ub{p}) \left[ b_{j,s}(\ub{q})
       V^\mu_{ij,2s}(\ub{p},\ub{q})\right.\right.\nonumber \\
      &&\left.\left.\rule{0.5in}{0in}
+ b_{j,-s}(\ub{q})
      U^\mu_{ij,-2s}(\ub{p},\ub{q})\right] 
            \sqrt{\xi_l}a_{l\mu}^\dagger(\ub{q}-\ub{p})
                    + H.c.\right\} ,  \nonumber
\eea
with $\epsilon^\mu = (-1,1,1,1)$ and the vertex functions given in \cite{SecDep}.

We work in a frame where the total transverse
momentum $\vec P_\perp$ is zero and 
expand the eigenfunction for the dressed-fermion state
with total $J_z=\pm \frac12$ in a Fock basis as
\bea \label{eq:FockExpansion}
\lefteqn{|\psi^\pm(\ub{P})\rangle=\sum_i z_i b_{i\pm}^\dagger(\ub{P})|0\rangle
  +\sum_{ijs\mu}\int d\ub{k} C_{ijs}^{\mu\pm}(\ub{k})b_{is}^\dagger(\ub{P}-\ub{k})
                                       a_{j\mu}^\dagger(\ub{k})|0\rangle}&& \\
 && +\sum_{ijks\mu\nu}\int d\ub{k_1} d\ub{k_2} C_{ijks}^{\mu\nu\pm}(\ub{k_1},\ub{k_2})
       \frac{1}{\sqrt{1+\delta_{jk}\delta_{\mu\nu}}}  
              b_{is}^\dagger(\ub{P}-\ub{k_1}-\ub{k_2})
                 a_{j\mu}^\dagger(\ub{k_1})a_{k\nu}^\dagger(\ub{k_2})|0\rangle, 
     \nonumber
\eea
where we have truncated the expansion to include at most two photons.
The $z_i$ are the amplitudes for the bare electron states, with $i=0$ for
the physical electron and $i=1$ for the PV electron.  The $C_{ijs}^{\mu\pm}$ are
the two-body wave functions for Fock states with an electron of flavor $i$
and spin component $s$ and a photon of flavor $j=0$, 1 or 2 and field component $\mu$,
expressed as functions of the photon momentum.  The upper index of $\pm$
refers to the $J_z$ value of $\pm\frac12$ for the eigenstate.
Similarly, the $C_{ijks}^{\mu\nu\pm}$ are the three-body wave functions
for the states with one electron and two photons, with flavors $j$ and $k$
and field components $\mu$ and $\nu$.

The Fock expansion is an eigenstate of the light-front Hamiltonian ${\cal P}^-$
with eigenvalue \\
$M^2/P^+$. The wave functions then satisfy the following
coupled integral equations:
\bea \label{eq:firstcoupledequation}
[M^2-m_i^2]z_i & = & \int d\ub{q} \sum_{j,l,\mu}\sqrt{\xi_l}(-1)^{j+l}\epsilon^\mu P^+
  \left[V_{ji\pm}^{\mu*}(\ub{P}-\ub{q},\ub{P})C_{jl\pm}^{\mu\pm}(\ub{q}) \right. \\
   &&  \rule{1in}{0mm} \left.   
     +U_{ji\pm}^{\mu*}(\ub{P}-\ub{q},\ub{P}) C_{jl\mp}^{\mu\pm}(\ub{q})\right], \nonumber
\eea
\bea \label{eq:secondcoupledequation}
\lefteqn{\left[M^2 - \frac{m_i^2 + q_\perp^2}{(1-y)} - \frac{\mu_l^2 + q_\perp^2}{y}\right]
  C_{ils}^{\mu\pm}(\ub{q}) }&& \\
    &=& \sqrt{\xi_l}\sum_j (-1)^j z_j P^+ 
      \left[\delta_{s,\pm 1/2}V_{ijs}^\mu(\ub{P}-\ub{q},\ub{P})
             +\delta_{s,\mp 1/2}U_{ij,-s}^\mu(\ub{P}-\ub{q},\ub{P})\right] \nonumber \\
    && +\sum_{ab\nu}(-1)^{a+b}\epsilon^\nu\int d\ub{q}' 
          \frac{2\sqrt{\xi_b}}{\sqrt{1+\delta_{bl}\delta^{\mu\nu}}}
    \left[V_{ais}^{\nu *}(\ub{P}-\ub{q}'-\ub{q},\ub{P}-\ub{q}')
             C_{abls}^{\nu\mu\pm}(\ub{q}',\ub{q}) \right. \nonumber \\
    &&   \rule{2in}{0mm} \left.  +U_{ais}^{\nu *}(\ub{P}-\ub{q}'-\ub{q},\ub{P}-\ub{q}')
             C_{abl,-s}^{\nu\mu\pm}(\ub{q}',\ub{q}) \right] , \nonumber
\eea
\bea \label{eq:thirdcoupledequation}
\lefteqn{\left[M^2 - \frac{m_i^2 + (\vec q_{1\perp}+\vec q_{2\perp})^2}{(1-y_1-y_2)} 
      - \frac{\mu_j^2 + q_{1\perp}^2}{y_1}  - \frac{\mu_l^2 + q_{2\perp}^2}{y_2}\right]
  C_{ijls}^{\mu\nu\pm}(\ub{q}_1,\ub{q}_2) }&& \\
    &=& \frac{\sqrt{1+\delta_{jl}\delta^{\mu\nu}}}{2}\sum_a (-1)^a \left\{
        \sqrt{\xi_j}\left[V_{ias}^\mu(\ub{P}-\ub{q}_1-\ub{q}_2,\ub{P}-\ub{q}_2)
                              C_{als}^{\nu\pm}(\ub{q}_2)  \right.  \right.  \nonumber \\
     && \rule{2in}{0mm}  \left.   +U_{ia,-s}^\mu(\ub{P}-\ub{q}_1-\ub{q}_2,\ub{P}-\ub{q}_2)
                              C_{al,-s}^{\nu\pm}(\ub{q}_2)\right]  \nonumber \\
      &&\rule{1in}{0mm}  
        + \sqrt{\xi_l}\left[V_{ias}^\nu(\ub{P}-\ub{q}_1-\ub{q}_2,\ub{P}-\ub{q}_1)
                              C_{ajs}^{\mu\pm}(\ub{q}_1)   \right. \nonumber \\
     &&  \left.\left. \rule{2in}{0mm}
                   +U_{ia,-s}^\nu(\ub{P}-\ub{q}_1-\ub{q}_2,\ub{P}-\ub{q}_1)
                              C_{aj,-s}^{\mu\pm}(\ub{q}_1)\right] \right\}. \nonumber
\eea

The anomalous moment $a_e$ can be computed from the spin-flip matrix element
of the electromagnetic current $J^+$~\cite{BrodskyDrell}.
At zero momentum transfer, we have $a_e=F_2(0)$ and
\bea  \label{eq:TwoPhotonae}
a_e&=&m_e\sum_{s\mu}\int d\ub{k}\epsilon^\mu \sum_{j=0,2}\xi_j
  \left(\sum_{i'=0}^1\sum_{k'=j/2}^{j/2+1}
    \frac{(-1)^{i'+k'}}{\sqrt{\xi_{k'}}}C_{i'k's}^{\mu+}(\ub{k})\right)^* \\
  && \times y\left(\frac{\partial}{\partial k_x}+i\frac{\partial}{\partial k_y}\right)
  \left(\sum_{i=0}^1\sum_{k=j/2}^{j/2+1}
    \frac{(-1)^{i+k}}{\sqrt{\xi_k}}C_{iks}^{\mu-}(\ub{k})\right) \nonumber \\
  && +m_e \sum_{s\mu\nu}\int d\ub{k_1} d\ub{k_2} \sum_{j,k=0,2} \xi_j\xi_k  \nonumber \\
  && \rule{1in}{0in} \times
            \left( \sum_{i'=0}^1 
                \sum_{l'=j/2}^{j/2+1} \sum_{m'=k/2}^{k/2+1} 
                  \frac{(-1)^{i'+l'+m'}}{\sqrt{\xi_{l'}\xi_{m'}}}
              \frac{\sqrt{2}C_{i'l'm's}^{\mu\nu+}(\ub{k_1},\ub{k_2})}
                          {\sqrt{1+\delta_{l'm'}\delta_{\mu\nu}}}\right)^* \nonumber \\
   && \times \sum_a \left[y_a \left(\frac{\partial}{\partial k_{ax}}
                                   +i\frac{\partial}{\partial k_{ay}}\right)\right]
                 \left(\sum_{i=0}^1 
                \sum_{l=j/2}^{j/2+1} \sum_{m=k/2}^{k/2+1} 
                  \frac{(-1)^{i+l+m}}{\sqrt{\xi_l\xi_m}}
              \frac{\sqrt{2}C_{ilms}^{\mu\nu-}(\ub{k_1},\ub{k_2})}
                          {\sqrt{1+\delta_{lm}\delta_{\mu\nu}}}\right) .
                          \nonumber
\eea
The terms that depend on the three-body wave functions $C_{ilms}^{\mu\nu\pm}$
are higher order in $\alpha$ than the leading two-body terms.  Given the numerical 
errors in the leading terms, these three-body contributions are not significant
and are not evaluated.  The important three-body contributions come from the
couplings of the three-body wave functions that will enter the calculation 
of the two-body wave functions.

\section{Solution of the Coupled Equations}

The first and third equations of the coupled system,
(\ref{eq:firstcoupledequation}) and (\ref{eq:thirdcoupledequation}),
can be solved for the bare-electron
amplitudes and one-electron/two-photon wave functions,
respectively, in terms of the one-electron/one-photon
wave functions. Substitution of these solutions into
the second integral equation (\ref{eq:secondcoupledequation})
yields a reduced integral
eigenvalue problem in the one-electron/one-photon sector:
\bea \label{eq:ReducedEqn}
\lefteqn{\left[M^2
  -\frac{m_i^2+q_\perp^2}{1-y}-\frac{\mu_j^2+q_\perp^2}{y}\right]
C_{ijs}^{\mu\pm}(y,q_\perp)=
\frac{\alpha}{2\pi}\sum_{i'}\frac{I_{iji'}(y,q_\perp)}{1-y}C_{i'js}^{\mu\pm}(y,q_\perp)}&& \\
 && \rule{1in}{0mm}
  +\frac{\alpha}{2\pi}\sum_{i'j's'\nu}\epsilon^\nu\int_0^1dy'dq_\perp^{\prime 2}
   J_{ijs,i'j's'}^{(0)\mu\nu}(y,q_\perp;y',q'_\perp)C_{i'j's'}^{\nu\pm}(y',q'_\perp) \nonumber \\
   &&  \rule{1in}{0mm}
   +\frac{\alpha}{2\pi}\sum_{i'j's'\nu}\epsilon^\nu\int_0^{1-y}dy'dq_\perp^{\prime 2}
   J_{ijs,i'j's'}^{(2)\mu\nu}(y,q_\perp;y',q'_\perp)C_{i'j's'}^{\nu\pm}(y',q'_\perp). \nonumber
\eea
There is a total of 48 coupled equations, with $i=0,1$; $j=0,1,2$; $s=\pm\frac12$;
and $\mu=\pm,(\pm)$.  

The first term on the right-hand side of (\ref{eq:ReducedEqn})
is the self-energy contribution~\cite{SecDep}:
\be  \label{eq:selfenergy}
I_{ili'}(y,q_\perp)
=\sum_{a,b}(-1)^{i'+a+b}\xi_b\int_0^1\frac{dx}{x}\frac{d^2k_\perp}{\pi}
\frac{m_i m_{i'} -2 \frac{m_i+m_{i'}}{1-x}m_a+\frac{m_a^2+k_\perp^2}{(1-x)^2}}
{\Lambda_l-\frac{m_a^2+k_\perp^2}{1-x}-\frac{\mu_b^2+k_\perp^2}{x}} ,
\ee
with
\be \label{eq:Lambda}
\Lambda_l\equiv \mu_l^2+(1-y)M^2-\frac{\mu_l^2+q_\perp^2}{y}.
\ee
The kernels $J^{(0)}$ and $J^{(2)}$ in the second and third terms
correspond to interactions with zero or two photons in intermediate states.
Details of these kernels can be found in \cite{TwoPhotonQED} and \cite{Thesis}.  

The presence of the flavor changing self-energies,
the $I_{ili'}$ with $i\neq i'$, generates
a fermion flavor mixing of the two-body wave functions~\cite{SecDep}.
To resolve this, we write the integral equations for these wave functions
in the form
\be \label{eq:MixingEquations}
A_{0j}C_{0js}^{\mu\pm} - B_jC_{1js}^{\mu\pm} = -\frac{\alpha}{2\pi}J_{0js}^{\mu\pm} , \;\;
B_j C_{0js}^{\mu\pm} + A_{1j} C_{1js}^{\mu\pm} = -\frac{\alpha}{2\pi} J_{1js}^{\mu\pm},
\ee
where $A_{ij}$ and $B_j$ are defined by
\be  \label{eq:Aij}
A_{ij}=\frac{m_i^2+q_\perp^2}{1-y}+\frac{\mu_j^2+q_\perp^2}{y}
    +\frac{\alpha}{2\pi}\frac{I_{iji}}{1-y}-M^2, \;\;
B_j=\frac{\alpha}{2\pi}\frac{I_{1j0}}{1-y}=-\frac{\alpha}{2\pi}\frac{I_{0j1}}{1-y},
\ee
and $J_{ijs}^{\mu\pm}$ is given by
\bea  \label{eq:Jmuijs}
J_{ijs}^{\mu\pm}&=&\sum_{i'j's'\nu}\epsilon^\nu\int_0^1dy'dq_\perp^{\prime 2}  
   J_{ijs,i'j's'}^{(0)\mu\nu}(y,q_\perp;y',q'_\perp)
            C_{i'j's'}^{\nu\pm}(y',q'_\perp) \\
  &&+\sum_{i'j's'\nu}\epsilon^\nu\int_0^{1-y}dy'dq_\perp^{\prime 2}
     J_{ijs,i'j's'}^{(2)\mu\nu}(y,q_\perp;y',q'_\perp)
            C_{i'j's'}^{\nu\pm}(y',q'_\perp) . \nonumber
\eea
We then construct wave functions that mix fermion flavors and
diagonalize the left-hand side of 
(\ref{eq:MixingEquations}):
$\tilde f_{ijs}^{\mu\pm}=A_{ij}C_{ijs}^{\mu\pm}+(-1)^i B_j C_{1-i,js}^{\mu\pm}$.
Solution of the resulting integral equations for the $f_{ijs}^{\mu\pm}$~\cite{TwoPhotonQED}
yields $\alpha$ as a function of $m_0$ and the PV masses.  Then for given
values of PV masses, we can seek the value of $m_0$ for which
$\alpha$ takes the standard physical value $e^2/4\pi$. 
The eigenproblem solution also yields the functions $\tilde f_{ijs}^{\mu\pm}$ which
determine the wave functions $C_{ijs}^{\mu\pm}$.  From these wave functions
we can compute physical quantities as expectation values with respect to
the projection~\cite{TwoPhotonQED} of the eigenstate onto the physical subspace.

The eigenvalue problem must first be solved for $M=0$, with the 
coupling strength parameter $\xi_2$ adjusted to yield $m_0=0$.  This
determines the value of $\xi_2$ that restores the chiral limit
nonperturbatively.  The eigenvalue problem can then be solved for
$M=m_e$, the physical mass of the electron, and the anomalous moment
calculated.

If we retain only the
self-energy contributions from the two-photon intermediate states,
the equations for the two-body wave functions become much simpler, and
the coupled integral equations can be reduced to the one-electron
sector.  There, they can be solved analytically, except for the
calculation of certain integrals~\cite{SecDep}.

\begin{figure}[t]
\centerline{\includegraphics[width=12cm]{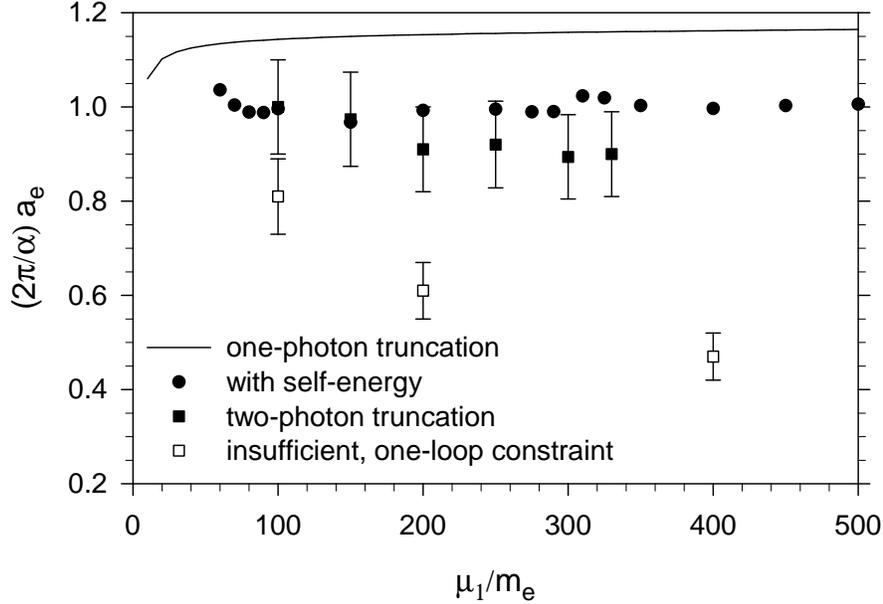}}
\caption{\label{fig:twophoton} The anomalous moment of the electron in
units of the Schwinger term ($\alpha/2\pi$) plotted versus
the PV photon mass, $\mu_1$, with the second PV photon mass, 
$\mu_2$, set to $\sqrt{2}\mu_1$
and the PV electron mass $m_1$ equal to $2\cdot10^4\,m_e$.
The solid squares are the result of the full two-photon truncation
with the correct, nonperturbative chiral constraint~\protect\cite{TwoPhotonQED}.
The open squares come from use of a perturbative, one-loop constraint.
Results for the one-photon truncation~\protect\cite{ChiralLimit}
(solid line) and the one-photon truncation with the
two-photon self-energy contribution~\protect\cite{SecDep} (filled circles)
are included for comparison.
The resolutions used for the two-photon results are $K=50$ to 150,
combined with extrapolation to $K=\infty$, and $N_\perp=20$.}
\end{figure}

\section{Results}

From the solutions to the eigenvalue problems, we compute the 
anomalous moment at fixed PV masses and fixed numerical
resolution.  We then study the behavior first as a function
of the numerical resolution, which requires extrapolation,
and then as a function of PV masses.  The numerical
resolution is marked by two parameters, $K$ and $N_\perp$, which
control the number of quadrature points used in the longitudinal
and transverse directions.  The numerical convergence
and extrapolation are illustrated in \cite{TwoPhotonQED}.

The results of the extrapolations are plotted in Fig.~\ref{fig:twophoton}.
Each value is close to the standard Schwinger result of
$\alpha/2\pi$ and independent of $\mu_1$, to within numerical error.
The results with only the two-photon self-energy contribution
are actually better than the full two-photon results.  This
discrepancy should be due to the absence of electron-positron
contributions, which are of the same order in $\alpha$ as
the two-photon contributions; without the electron-positron
contributions, we lack the cancellations that typically take place
between contributions of the same order.

We also see that the inclusion of the self-energy contribution
is a significant improvement over the one-photon truncation.
Thus, we expect that inclusion of three-photon self-energy effects
will improve the two-photon results.

Figure~\ref{fig:twophoton} also includes results obtained for the
two-photon truncation when only the one-loop chiral constraint
is satisfied.  Without the full
nonperturbative constraint, the results are very sensitive
to the PV photon mass $\mu_1$.  This behavior repeats the pattern observed
in \cite{ChiralLimit} for a one-photon truncation without the
corresponding one-loop constraint.  The resulting $\mu_1$
dependence is illustrated in Fig.~2 of \cite{ChiralLimit}.  Thus,
a successful calculation requires that the symmetry of the chiral
limit be maintained.

\acknowledgments
The work reported here was done in collaboration with 
J.R. Hiller
and supported in part by the Minnesota Supercomputing Institute.


\begin{thebibliography}{99}

\bibitem{Lattice} For reviews of lattice theory, see 
   M. Creutz, L. Jacobs and C. Rebbi,
   \emph{Phys.\ Rep.}\ {\bf 95} (1983) 201;
   J.B. Kogut, \emph{Rev.\ Mod.\ Phys.}\ {\bf 55} (1983) 775;
   I. Montvay, {\em ibid}.\ {\bf 59} (1987) 263;
   A.S. Kronfeld and P.B. Mackenzie,
   \emph{Ann.\ Rev.\ Nucl.\ Part.\ Sci.}\ {\bf 43} (1993) 793;
   J.W. Negele, \emph{Nucl.\ Phys.}\ {\bf A553} (1993) 47c;
   K.G.~Wilson,
   \emph{Nucl.\ Phys.}\ B (Proc.\ Suppl.) {\bf 140} (2005) 3;
   J.M. Zanotti, 
   \pos{PoS(LAT2008)007}.
   For recent discussions of meson properties and charm physics, see for example
   C. McNeile and C. Michael [UKQCD Collaboration], 
   \emph{Phys.\ Rev.}\ D {\bf 74} (2006) 014508;
   I.~Allison {\em et al.}  [HPQCD Collaboration],
   \emph{Phys.\ Rev.}\  D {\bf 78} (2008) 054513.  
   
\bibitem{TransLattice} M.~Burkardt and S.~Dalley,
   \emph{Prog.\ Part.\ Nucl.\ Phys.}\ {\bf 48} (2002) 317 and references therein; 
   S.~Dalley and B.~van~de~Sande, 
   \emph{Phys.\ Rev.}\ D {\bf 67} (2003) 114507;
   D. Chakrabarti, A.K. De, and A. Harindranath,
   \emph{Phys.\ Rev.}\ D {\bf 67} (2003) 076004;
   M. Harada and S. Pinsky, 
   \emph{Phys.\ Lett.}\ B {\bf 567} (2003) 277;
   S.~Dalley and B.~van de Sande,
   \emph{Phys.\ Rev.\ Lett.}\  {\bf 95} (2005) 162001; 
   J.~Bratt, S.~Dalley, B.~van de Sande, and E.~M.~Watson,
   \emph{Phys.\ Rev.}\ D {\bf 70} (2004) 114502. 
   For work on a complete light-cone lattice, see
   C. Destri and H.J. de Vega,
   \emph{Nucl.\ Phys.}\ {\bf B290} (1987) 363;
   D. Mustaki, \emph{Phys.\ Rev.}\ D {\bf 38} (1988) 1260.

\bibitem{DSE} C.D. Roberts and A.G. Williams,
   \emph{Prog.\ Part.\ Nucl.\ Phys.}\ {\bf 33} (1994) 477;
   P. Maris and C.D. Roberts, \emph{Int.\ J. Mod.\ Phys.}\ {\bf E12} (2003) 297;
   P.C. Tandy, \emph{Nucl.\ Phys.}\ B (Proc.\ Suppl.) {\bf 141} (2005) 9.
   
\bibitem{Glazek}  S.~D.~Glazek and R.~J.~Perry,
  \emph{Phys.\ Rev.}\ D {\bf 78} (2008) 045011;
  S.D.~G{\l}azek and J.~Mlynik,
  \emph{Phys.\ Rev.}\ D {\bf 74} (2006) 105015; 
  S.D.~G{\l}azek,
  \emph{Phys.\ Rev.}\ D {\bf 69} (2004) 065002; 
  S.D.~G{\l}azek and J.~Mlynik,
  \emph{Phys.\ Rev.}\ D {\bf 67} (2003) 045001;
  S.D.~G{\l}azek and M.~Wieckowski,
  \emph{Phys.\ Rev.}\ D {\bf 66} (2002) 016001.

\bibitem{bhm} S.J. Brodsky, J.R. Hiller, and G. McCartor,
  \emph{Phys.\ Rev.}\ D {\bf 58} (1998) 025005;
%
  {\bf 60} (1999) 054506;
%
  {\bf 64} (2001) 114023; 
%
  \emph{Ann.\ Phys.}\ {\bf 296} (2002) 406; 
%
{\bf 305} (2003) 266; 
%
{\bf 321} (2006) 1240; 
%
S.J. Brodsky, V.A. Franke, J.R. Hiller, G. McCartor, 
S.A. Paston, and E.V. Prokhvatilov,
\emph{Nucl.\ Phys.}\ B {\bf 703} (2004) 333. 

\bibitem{SectorDependent} {R.J. Perry, A. Harindranath, and K.G. Wilson,
   \emph{Phys.\ Rev.\ Lett.}\ {\bf 65} (1990) 2959;
   R.J. Perry and A. Harindranath,
   \emph{Phys.\ Rev.}\ D {\bf 43} (1991) 4051.}

\bibitem{Karmanov}
  V.~A.~Karmanov, J.~F.~Mathiot, and A.~V.~Smirnov,
  \emph{Phys.\ Rev.}\ D {\bf 77} (2008) 085028;
  arXiv:1006.5640 [hep-th].

\bibitem{Vary} J.P. Vary {\em et al}.,
   \emph{Phys.\ Rev.}\ C {\bf 81} (2010) 035205.

\bibitem{PauliVillars} W. Pauli and F. Villars,
   \emph{Rev.\ Mod.\ Phys.}\ {\bf 21} (1949) 434.

\bibitem{ChiralLimit} S.S. Chabysheva and J.R. Hiller,
   \emph{Phys.\ Rev.}\ D {\bf 79} (2009) 114017.

\bibitem{Thesis} S.S. Chabysheva,
   \emph{A nonperturbative calculation of the electron's anomalous magnetic moment},
   Ph.D. thesis, Southern Methodist University 
   ProQuest Dissertations \& Theses 3369009 2009.
   
\bibitem{SecDep} S.S. Chabysheva and J.R. Hiller,
\emph{Ann.\ Phys.}\ {\bf 325} (2010) 2435.

\bibitem{TwoPhotonQED} S.S. Chabysheva and J.R. Hiller,
   \emph{Phys.\ Rev.}\ D {\bf 81} (2010) 074030.
   
\bibitem{Dirac} {P.A.M. Dirac, 
   \emph{Rev.\ Mod.\ Phys.}\ {\bf 21} (1949) 392.}

\bibitem{DLCQreview} For reviews of light-cone quantization, see
   M. Burkardt, \emph{Adv.\ Nucl.\ Phys.}\ {\bf 23}, 1 (2002);
   S.J. Brodsky, H.-C. Pauli, and S.S. Pinsky, 
   \emph{Phys.\ Rep.}\ {\bf 301} (1998) 299. 
   
\bibitem{BrodskyDrell} S.J. Brodsky and S.D. Drell,
\emph{Phys.\ Rev.}\ D {\bf 22} (1980) 2236.

\end{thebibliography}
\end{document}